\documentclass{./styles/svproc}
\usepackage{url}
\usepackage{graphicx}
\usepackage{multirow}
\usepackage{multicol}
\usepackage{cellspace}
\usepackage{booktabs}
\usepackage{xcolor}

\begin{document}
\mainmatter              
%
\title{Decentralized Healthcare Systems with Federated Learning and Blockchain}

%
%
\author{Abdulrezzak Zekiye  \and Öznur Özkasap}
%
%
%
\institute{Department of Computer Engineering, Koç University, İstanbul, Türkiye\\
\email{\{azakieh22, oozkasap\}@ku.edu.tr}
}

\maketitle              

\begin{abstract}
Artificial intelligence (AI) and deep learning techniques have gained significant attraction in recent years, owing to their remarkable capability of achieving high performance across a broad range of applications. However, a crucial challenge in training such models is the acquisition of vast amounts of data, which is often limited in fields like healthcare. In this domain, medical data is typically scattered across various sources such as hospitals, clinics, and wearable devices. The aggregated data collected from multiple sources in the healthcare domain is sufficient for training advanced deep learning models. However, these sources are frequently hesitant to share such data due to privacy considerations. To address this challenge, researchers have proposed the integration of blockchain and federated learning to develop a system that facilitates the secure sharing of medical records. This work provides a succinct review of the current state of the art in the use of blockchain and federated learning in the decentralized healthcare domain.

\end{abstract}
\section{Introduction}
The growing collection of data has raised serious concerns regarding privacy, particularly for medical data, which can be obtained from healthcare providers and wearable devices. Such data encompasses a wide range of information, including treatments, drugs, prescriptions, tests, images, and vital signs. Nevertheless, medical data is often dispersed across multiple entities, and the availability of large-scale medical datasets is limited. For instance, researchers in \cite{Hayyolalam:Hybrid} trained a machine learning model with only 303 records. To tackle these challenges, the combination of Federated Learning (FL) and Blockchain (BC) technology has been proposed to enable data sharing while preserving privacy. By means of federated learning, entities train a global machine learning model and then share the parameters of the model instead of raw data. Blockchain, generally, can be used to make the system decentralized, transparent, immutable, and self-controlled by using smart contracts. 

There exist reviews about using blockchain in the medical fields as given in \cite{Sai:Confluence} and \cite{Hashim:Precision}. The usage of federated learning for healthcare is discussed in \cite{Rahman:Federated}. To our knowledge, there is no review of the usage of both techniques in an integrated manner in the field of medical systems. The contribution of this paper is providing a review of state-of-the-art works utilizing both federated learning and blockchain in the medical field. In the rest of this extended abstract, we present some of the latest research that used both federated learning and blockchain in the medical field with a graphical representation of blockchain and federated learning-enabled systems.

\section{State-of-the-art: Blockchain and Federated Learning Empowered Healthcare Solutions}
\begin{figure}[!ht]
	\centering
  \includegraphics[width=0.8\linewidth]{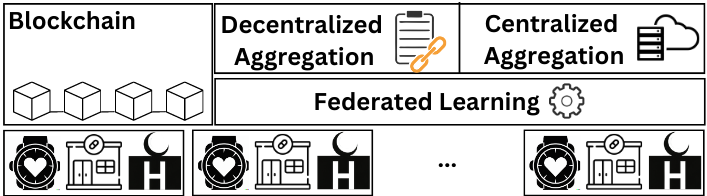}

    \caption{A general representation of blockchain and federated learning-enabled systems}
    \label{fig:systemmodel}
\end{figure}

This section presents an overview of the representative recent studies that employed blockchain and federated learning in the healthcare domain. The type of used blockchain, the type of aggregation performed in federated learning, the proposed solution and the application types are summarized in Table \ref{table:summary}. 
A graphical representation of the integration of blockchain and federated learning in healthcare systems is presented in Figure \ref{fig:systemmodel}. The diagram displays entities, including hospitals, clinics, and wearable devices, as the primary sources of data. The federated learning component trains local models on local data, either through on-site processing or by sending the data to nodes with sufficient computational power. Model aggregation can be performed either through a centralized server or through a decentralized mechanism facilitated by a blockchain-based smart contract.

Researchers in \cite{Rahman:Secure} utilized blockchain and federated learning to build a secure and provenance framework to be used in the Internet of Health Things (IoHT). They tested the built framework with COVID-19 detection scenarios. Federated learning has been used to train data locally instead of sharing it. Differential privacy was applied on top of the federated learning to increase privacy protection. The role of blockchain in their framework is managing the federated learning process in a decentralized way. Specifically, the blockchain had a role in aggregating the global model, handling clients’ reputations and the quality of their contributions, and storing the local and global models in Interplanetary File System (IPFS). In addition, the blockchain in this solution allows users to track the decision path, i.e. the used model and dataset.

\setlength\cellspacetoplimit{2pt}
\setlength\cellspacebottomlimit{2pt}
\begin{table}
\caption{Summary and comparison of the state-of-the-art solutions}
\label{table:summary}
\begin{center}
\begin{tabular}{|Sc|Sc|Sc|Sc|Sc|}

\hline
\textbf{Ref} & \parbox[c]{3cm}{\textbf{Blockchain Type}} &   \textbf{Aggregation} & \textbf{Solution} & \textbf{Application Type}
\\ 
\hline
\cite{Rahman:Secure} & Permissioned  & Decentralized & \parbox[c]{3cm}{Blockchain-managed Federated Learning} & \parbox[c]{3cm}{Security and Provenance in IoHT}
\\ \hline
\cite{Lian:Blockchain} & Permissioned & Centralized & \parbox[c]{3cm}{ Blockchain-based two-stage platform }  & \parbox[c]{3cm}{FL for Non-IID Data in IoMT}
\\ \hline
\cite{Abou:When} & Public & Centralized &\parbox[c]{3cm}{Blockchain \& federated learning with encryption}  & \parbox[c]{3cm}{Privacy-preserving clinician collaboration}
\\ \hline
\cite{Shae:Transform} & Public & Not Mentioned & \parbox[c]{3cm}{Blockchain as computing architecture}  & \parbox[c]{3cm}{Building large medical datasets}

\\ \hline

\cite{Mazzocca:Federated} & Permissioned & Decentralized & \parbox[c]{3cm}{Risk-Based authorization middleware}  & \parbox[c]{3cm}{Access control to IoMT data}
\\ \hline
\end{tabular}
\end{center}
\end{table}

Using federated learning has a problem when dealing with non-independent
identically distributed (non-IID) data. The reason is that the data are different at each federated agent and there is a risk of training each local model using imbalanced data. Researchers in \cite{Lian:Blockchain} implemented a two-step solution to tackle this problem. The first step is building a uniformly distributed subset of the data by receiving raw data from the participants. The composed dataset is then sent to the participant to include in their training along with their local dataset. This way, the problem of having imbalanced data is less. The blockchain here is used to share the composed dataset. The second step is having each participant training the machine learning model using their own private data along with the shared dataset in the first step.

In \cite{Abou:When}, a blockchain and federated learning-based platform was built to preserve privacy and distribute the learning process between different clinics. The aggregation of local models has been done by centralized nodes using Secure Multiparty
Computation (SMPC) technique. The nodes encrypt the aggregated model to increase privacy and send the encrypted model to the leader, which is the entity that initialized the platform. The leader decrypts the received models, does aggregation, then sends the global model to the Software Defined Networks (SDNs) if further training is needed.
Simulation using a Breast Cancer Dataset has been done to test the proposed system.

The work presented in \cite{Shae:Transform} includes an approach to construct large medical datasets from diverse data sources. Blockchain technology is leveraged to regulate access to off-chain nodes. Additionally, federated learning is employed with modifications to train local models on locally stored data, thereby maintaining privacy through decentralized training.

A blockchain and federated learning middleware that is responsible for deciding whether to give an entity access to the patients' data or not was built in \cite{Mazzocca:Federated}. 
Federated learning role in their proposed solution is predicting the risk level depending on the patient's condition and the information of the entity requesting access to the data. The trained AI model will give three different levels of risk: critical, serious, and stable. According to those levels, the decision of giving access or not is taken.  
Blockchain was used to make the system decentralized, i.e. replacing the centralized server for aggregating the global model. The global and local models are stored on the blockchain as well.

\section{Conclusion}
Combining blockchain and federated learning has emerged as a promising solution for sharing medical data while preserving privacy. The decentralized nature of blockchain and the ability of federated learning to train models locally while sharing only model parameters make this approach well-suited for healthcare. The reviewed papers used both premissionless (public) and permissioned blockchains and aggregated local models either through a centralized server or decentralized smart contracts. Blockchain played a crucial role in decentralizing the system, managing privacy, and authorizing access to required data.

Remarkable results have been achieved by combining federated learning and blockchain techniques to classify data located at different nodes. According to \cite{Rahman:Secure}, a success rate of 85\% in classifying COVID-19 cases has been achieved using this approach. Furthermore, \cite{Lian:Blockchain} tested their system with three different datasets and demonstrated that sharing a small amount of raw data while conducting federated learning can increase accuracy when dealing with non-IID data. In another study, \cite{Abou:When} trained a deep learning model on the Breast Cancer Wisconsin dataset by distributing it over several nodes, achieving an f1-score of approximately 99\%. Finally, researchers in \cite{Mazzocca:Federated} compared their platform with a centralized approach and demonstrated that training a model on small datasets dispersed through different nodes can achieve results that are similar to those of a centralized solution with a large dataset.

\section{Acknowledgment}
This work was supported in part by TUBITAK 2247-A Award 121C338.
%
%
%

\end{document}